\documentclass[]{jpsj3}
\usepackage{graphicx}
\usepackage{dcolumn}
\usepackage{bm}

\def\gsim{\buildrel {\textstyle >}\over {_\sim}}
\def\lsim{\buildrel {\textstyle <}\over {_\sim}}

\title{Novel spin-liquid states in the frustrated Heisenberg antiferromagnet on the honeycomb lattice}

\author{Soichiro Okumura, Hikaru
Kawamura\thanks{E-mail:kawamura@ess.sci.osaka-u.ac.jp}, Tsuyoshi
Okubo, and Yukitoshi Motome$^{1}$}

\inst{Department of Earth and Space Science, Faculty of Science, Osaka University, Toyonaka 560-0043, Japan\\
$^{1}$Department of Applied Physics, University of Tokyo, Tokyo 113-8656, Japan}

\abst{Recent experiment on a honeycomb-lattice Heisenberg antiferromagnet
(AF) Bi$_3$Mn$_4$O$_{12}$(NO$_3$) revealed a novel spin-liquid-like
behavior down to low temperature, which was ascribed to the
frustration effect due to the competition between the AF nearest- and
next-nearest-neighbor interactions $J_1$ and $J_2$. Motivated by the
experiment, we study the ordering of the $J_1$ -$J_2$ frustrated
classical Heisenberg AF on a honeycomb lattice both by a
low-temperature expansion and a Monte Carlo simulation. The model has
been known to possess a massive degeneracy of the ground state, which,
however, might be lifted due to thermal fluctuations leading to a
unique ordered state, the effect known as `order-by-disorder'.  We
find that the model exhibits an intriguing ordering behavior,
particularly near the AF phase boundary. The energy scale of the
order-by-disorder is suppressed there down to extremely low
temperatures, giving rise to exotic spin-liquid states like a
``ring-liquid'' or a ``pancake-liquid'' state accompanied by the
characteristic spin structure factor and the field-induced
antiferromagnetism. We argue that the recent experimental data are
explicable if the system is in such exotic spin-liquid states.
}

\kword{frustration, honeycomb lattice, spin liquid, order by disorder}

\begin{document}
\maketitle
\section{Introduction}
 Recently, growing interest arises in the ordering of geometrically
 frustrated magnets. Particular attention has been paid to possible
 `spin-liquid' states stabilized due to the frustration effect,  where
 spins remain to be disordered down to very low temperature without
 showing the standard magnetic long-range order
 \cite{Anderson}. Usually, geometrical frustration is realized in
 triangle-based lattices like the triangular, kagome and pyrochlore
 lattices combined with the antiferromagnetic (AF) coupling. By
 contrast, the honeycomb lattice, a hexagon-based lattice shown in
 Fig.1, is bipartite and is usually regarded as an unfrustrated
 lattice since it can accommodate the standard AF `up-down'
 order. However, if the AF interaction works between the
 next-nearest-neighbor (nnn) sites  in addition to the nearest-neighbor
 (nn) sites, the frustration effect arises due to the competition
 between the nn and the nnn couplings $J_1$ and $J_2$: See
 Fig.1. Since the honeycomb lattice is a loosely-coupled lattice with
 the number of the nn sites only three, it might be susceptible to the
 fluctuation effect caused by frustration, and its ordering property is
 of special interest. Quantum magnetism on the honeycomb lattice has attracted much attention \cite{Einarsson,Takano,Kitaev,Kivelson,Cabra}. Interest in the honeycomb lattice is also further promoted by the recent upsurging research interest in graphenes \cite{Kane,Rachel}. 
\begin{figure}
 \begin{center}
 \includegraphics[width=6cm]{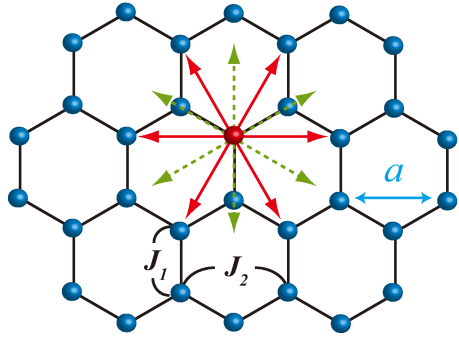}
  \end{center}
 \caption{(Color online) A honeycomb lattice, where $J_1$ and $J_2$ represent the
 nearest-neighbor (nn) and the next-nearest-neighbor (nnn) interactions,
 and $a$ is the lattice spacing of the triangular Bravais lattice equal
 to the nnn distance of the honeycomb lattice. The green (red) arrows
 represent six equivalent nn (nnn) directions. }
\end{figure}

 An earlier theoretical study by Katsura {\it et al\/} on the honeycomb-lattice classical Heisenberg model with the nn and the nnn AF couplings ($J_1$-$J_2$ model) revealed that, when the nnn AF coupling is moderately strong $J_2/J_1 > 1/6$, the classical ground state is infinitely degenerate, {\it i.e.\/}, the ground-state manifold of the model consists of a set of spiral states characterized by generally incommensurate wavevectors $\bm{q}$ which form the ring surrounding the AF point in the wavevector space, rather than the discrete points \cite{Katsura}. If the nnn coupling is weaker $J_2/J_1\leq 1/6$, on the other hand, the energy minimum occurs at the AF point. Such features are demonstrated in the Fourier-transformed energy of the model shown in Fig.2(a) and (b). Based on this observation, Katsura {\it et al\/} suggested that the ground state of the model might be disordered  at $J_2/J_1 > 1/6$ because the system might continue to fluctuate over these degenerate states characterized by different $\bm{q}$.
\begin{figure}
\begin{center}
 \includegraphics[width=8cm]{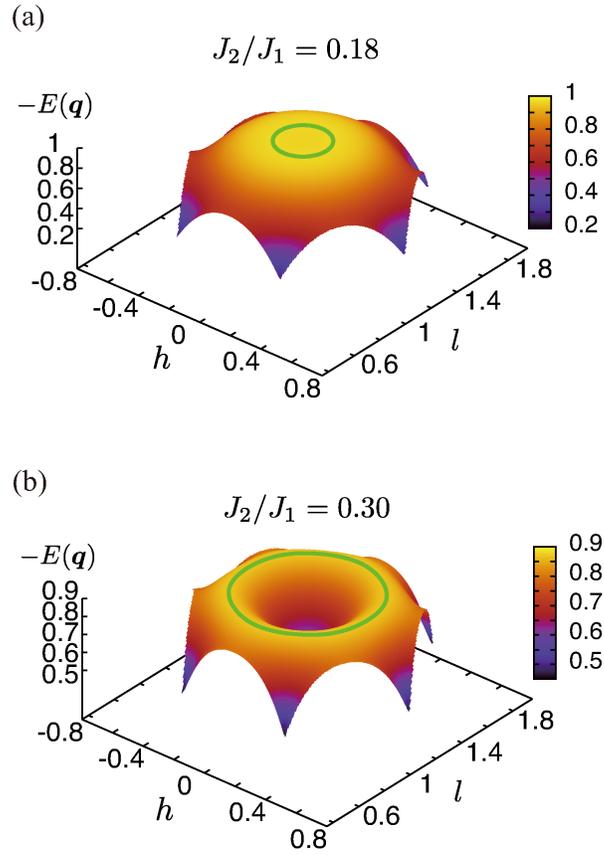}
\end{center}
 \caption{(Color online) The minus of the Fourier-transformed energy of the model is plotted in the $(h,l)$ plane,  where $\bm{q} = \frac{2\pi}{a}(h,l)$ is the wavevector, for the case of $J_2/J_1=0.18$ (a), and  of $J_2/J_1=0.30$ (b). The green ring represents the degenerate line of the ground state, which appears surrounding the AF point ($0,2/\sqrt{3}$). }
\end{figure}

 Interest in the honeycomb AF is promoted by the recent experiment by Azuma {\it et al\/} who observed that the $S=3/2$ honeycomb-lattice Heisenberg AF Bi$_3$Mn$_4$O$_{12}$(NO$_3$) (BMNO) exhibited a spin-liquid-like behavior down to low temperature 0.4K, much lower than the Curie-Weiss temperature $T_{CW}\simeq -257$K \cite{Azuma,ESR,Matsuda}. The frustration effect due to the nnn coupling was invoked to explain the observed spin-liquid-like behavior. It thus remains most interesting to understand the ordering process of the frustrated honeycomb-lattice Heisenberg AF.

 In fact, a naive expectation linking the observed infinite degeneracy to the disordered ground state needs careful examination, since such degeneracy is often lifted by fluctuations, either thermal or quantum, a phenomenon known as `order-by-disorder' \cite{Villain}, which is often observed in frustrated spin systems \cite{Kawamura,Henley,Balents}. In the honeycomb-lattice Heisenberg AF, one naturally expects that an infinite ring-like degeneracy of the ground state might be lifted by fluctuations,  leading to a unique spiral ordered state characterized by a unique $\bm{q}$. The candidate directions selected might be either the nn or the nnn directions as shown in Fig.1.

 We find that such an order-by-disorder mechanism is certainly operative in the honeycomb-lattice Heisenberg AF, leading to a unique ordered state where the threefold [$Z_3$] directional symmetry of the honeycomb lattice is spontaneously broken. Usually, the order-by-disorder occurs at the energy scale of the main coupling, {\it i.e.\/}, at the order of the Curie-Weiss temperature $T_{CW}$ or the nn coupling $J_1$, if somewhat suppressed due to the frustration effect. Therefore, the transition temperature to a unique ordered state selected by the order-by-disorder is not extremely low. An example might be a field-induced plateau phase realized in the Heisenberg AF on the triangular and kagome lattices \cite{KM85,Zhitomirsky}. Emergence of the spin-liquid naively expected from the ground-state degeneracy might thereby be hampered in reality. Such limitations might particularly be severe in three-dimensional (3D) frustrated magnets.

 In such circumstances, an interesting possibility arises in the 2D honeycomb-lattice Heisenberg AF. We observe that, in this model near its AF phase boundary $J_2/J_1 \gsim 1/6$, the energy scale associated with the order-by-disorder is determined by $J_2-\frac{1}{6}J_1$ rather than by $J_1$ or $J_2$ itself, becoming arbitrary small around the AF phase boundary. Indeed, it tunes out that this regime, with the help of the ring-like degeneracy, gives rise to a variety of interesting behaviors such as exotic types of spin-liquid states, which we call ``ring-liquid'' and the ``pancake-liquid'' states, as well as the field-induced AF order.

\section{Model and method}

 The model we consider is the classical Heisenberg AF on the 2D
 honeycomb lattice, whose Hamiltonian is given by
\begin{equation}
 \mathcal{H} = J_1 \sum_{\langle i,j\rangle} \bm{S}_i\cdot\bm{S}_j + J_2
 \sum_{\langle\langle i,j\rangle\rangle} \bm{S}_i\cdot\bm{S}_j,
\end{equation}
where the sum is taken over all nn and nnn pairs for $J_1$ and $J_2$, respectively. We study the ordering properties of the model both by a low-temperature expansion and by a  Monte Carlo (MC) simulation.

 The low-temperature expansion is made following the method of
 Ref.\cite{Balents}, expanding around an arbitrary state in the
 ground-state manifold at the harmonic order. Some of the details is
 given in Appendix.

 MC is performed on the basis of the standard heat-bath method combined with the over-relaxation technique and the temperature-exchange method. Various values of $j_2=J_2/J_1$ in the range $j_2=$ [0.17, 0.5] are studied, employing both periodic and free boundary conditions (BCs). The lattice contains $2\times L^2$ spins, with $L$ ranging between $24\leq L\leq 72$. In the present paper, we study mainly the case of $j_2= 0.1830 \cdots $, which corresponds to $1/[4\cos \frac{q}{2}+2]$ with $q=\frac{4\pi}{12}$, for systems under periodic BCs. This particular value of $j_2$ is chosen to minimize the finite-size effect due to the mismatch between the incommensurability of the helix and the applied periodic BCs: The helix along the nnn direction becomes commensurate  with the lattice at this value of $j_2$ if the $L$-value is chosen as multiples of twelve.

\section{Results}
First, we report on our results of the low-temperature expansion. We find that thermal fluctuations select a particular incommensurate spiral state whose helix axis runs along the nnn direction for $J_2/J_1\equiv j_2\lsim 0.232$, but along the nn direction for $j_2 \gsim 0.232$: See Fig.3. The calculated free-energy difference between these two directions, $\Delta F$, is shown in Fig.3 as a function of $J_2/J_1$. Over the range $1/6 < j_2 \lsim 0.232$, $\Delta F$ stays small, suggesting that the selection is weak, whereas, for $j_2 \gsim 0.232$, $\Delta F$ is large. Our result of thermal selection differs from that of the recent $T=0$ calculation of Ref.\cite{Mulder} for the quantum honeycomb $J_1$-$J_2$ model, where the nn direction was selected irrespective of the $j_2$-value for $1/6 < j_2 < 0.5$.
\begin{figure}
\begin{center}
\includegraphics{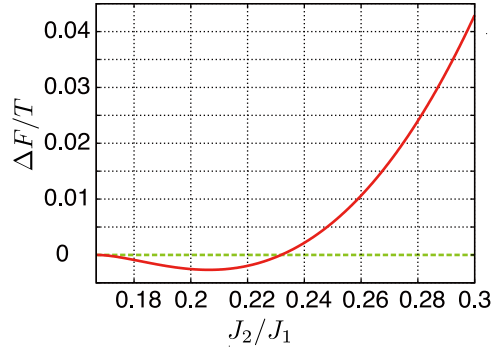}
\end{center}
 \caption{(Color online) The $J_2/J_1$-dependence of the free-energy difference between the spiral states running along the nn direction and along the nnn direction, $\Delta F \equiv F({\rm nnn}) - F({\rm nn})$, calculated by the low-temperature expansion. At $J_2/J_1 \lsim 0.232$ ($\gsim 0.232$), the nnn (nn) direction has the lower free energy, whereas the difference is small for $J_2/J_1 \lsim 0.232$.}
\end{figure}

 Next, we present our MC results. In Fig.4(a), we show
 the specific heat per spin for $j_2=0.1830$. (Here and below, the unit
 of the energy and the temperature is taken to be $J_1$.) The specific
 heat exhibits three peaks, a broad peak around $T\simeq 0.12$
 associated with the onset of the short-range order, a sharp diverging
 peak at  $T=T_c \simeq 0.013$ associated with a phase transition, and
 another peak at a lower temperature $T=T^*$ whose position and height
 depend on the system size $L$ considerably. The transition at $T_c$
 seems continuous. The Curie-Weiss temperature is estimated to be
 $T_{CW}\simeq -1.68$: See the inset of Fig.4(b). The diverging peak at
 $T_c\simeq 0.013$ is associated with the spontaneous breaking of the
 $Z_3$ directional symmetry of the lattice. 

 This can be confirmed from
 the order parameter $m_3$ describing the $Z_3$ directional-symmetry
 breaking. This order parameter is defined by 
\begin{equation}
m_3=\langle
 |\bm{m}_3|\rangle ,\ \ \ \ \bm{m}_3=\epsilon_1\hat e_1+\epsilon_2\hat
 e_2+\epsilon_3\hat e_3, 
\end{equation}
where $\hat e_1=(0,1)$, $\hat
 e_2=(-\frac{\sqrt{3}}{2},-\frac{1}{2})$ and $\hat
 e_3=(\frac{\sqrt{3}}{2},-\frac{1}{2})$, and $\epsilon_\mu$
 ($\mu=1,2,3$) is the total nn bond energy (normalized per bond) along
 the $\hat e_\mu$ direction.  When the spiral runs along the nnn
 direction, $m_3$ takes a value $1-\cos \frac{q}{2}$ at $T=0$. For
 $J_2/J_1=0.1830$, the saturation value is  then $m_3(T=0) =
 1-\frac{\sqrt 3}{2} \simeq 0.134 $. As can be seen from  Fig.4(b), the directional order sets in at $T_{c}\simeq 0.013$ which coincides with the specific-heat peak.  Even below $T_{c}$, $m_3$ does not saturate but increases further, and saturates below $T^*$.
\begin{figure}
\begin{center}
\includegraphics[width=8cm]{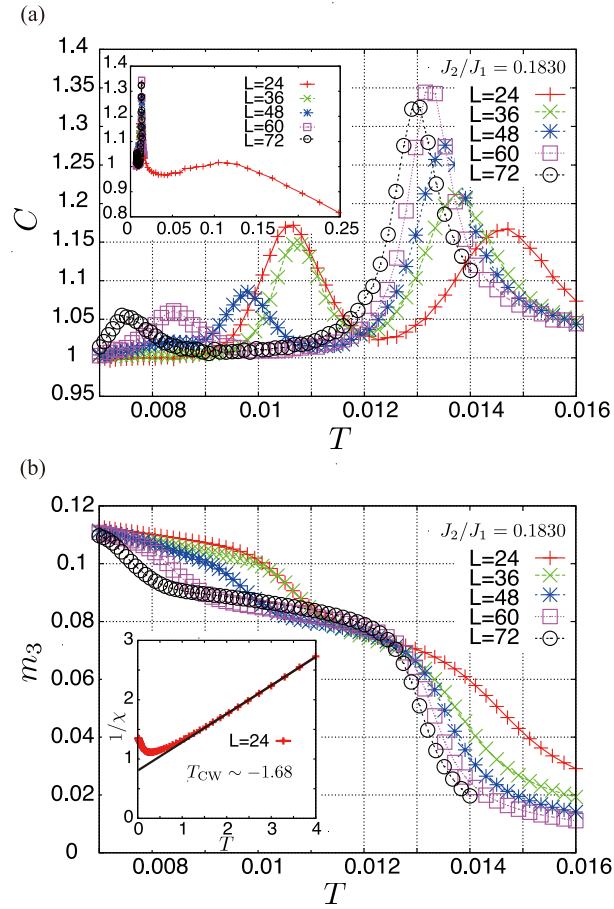}
\end{center}
 \caption{
(Color online) The temperature and size dependence of several physical quantities at $J_2/J_1=0.1830$. (a) The specific heat per spin. The inset represents a wider temperature range. (b) The threefold directional-symmetry-breaking order parameter $m_3$. The inset represents the Curie-Weiss plot, {\it i.e.\/}, the temperature dependence of the inverse susceptibility per spin. The Curie-Weiss temperature is estimated to be $T_{CW}\simeq -1.68$.
}
\end{figure}

 In Figs.5 and 6, we show the spin structure factor $F(\bm{q})$ in the ($h,l$) plane for the case of  $J_2/J_1=0.1830$, where $\bm{q}=\frac{2\pi}{a}(h,l)$ and $a$ is the lattice constant of the underlying Bravais lattice: See Fig.1. The spin structure factor is defined by $F(\bm{q})= \left\langle |\bm{S}_{\bm{q}}|^2 \right\rangle$, where $\langle \cdots \rangle$ represents the thermal average and $\bm{S}_{\bm{q}}$ is the Fourier transform of the spins. Fig.5 represents $F(\bm{q})$  at a temperature slightly above $T_{c}$ shown in the entire $\bm{q}$ space around the $\Gamma$ point, while Fig.6 is a magnified view of a part of the $\bm{q}$ space around an AF point $(0,2/\sqrt{3})$ at several temperatures across $T_{c}$ and $T^*$. At a temperature below $T^*$, $F(\bm{q})$ exhibits sharp point-like spots in the nnn direction, consistently with the prediction of the low-temperature expansion. In fact, the nnn direction is chosen at lower temperatures for $1/6\lsim j_2 \lsim 0.232$ for large enough lattices under periodic boundary conditions. However, the other nn direction tends to be stabilized for smaller sizes. Finite-size effects seem significant here, presumably reflecting the small free-energy difference between the two directions shown in Fig.3.

\begin{figure*}[th]
\begin{center}
 \includegraphics[width=8cm]{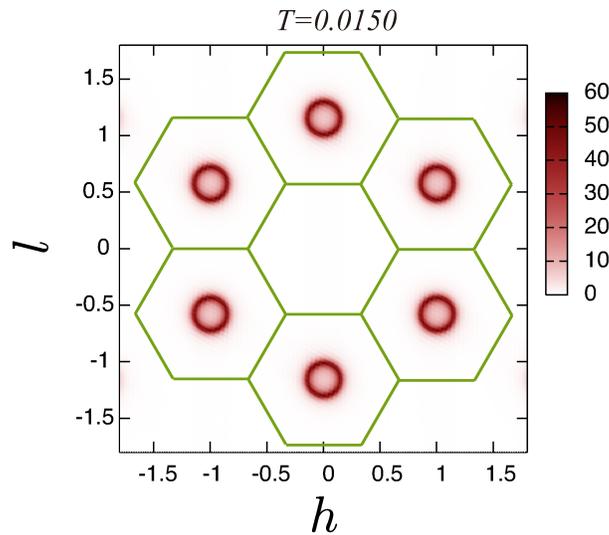}
\end{center}
 \caption{
(Color online) The intensity plot of the spin structure factor in the entire $(h,l)$ plane  around the $\Gamma $ point $(0,0)$ for the case of $J_2/J_1=0.1830$, where $\bm{q}=\frac{2\pi}{a}(h,l)$. The temperatures is $T=0.0150$ and the lattice size is $L=72$. In this simulation, we turn off the temperature-exchange process when measuring $F(\bm{q})$ to probe the way of the symmetry breaking clearly, while it is turned on in equilibrating the system prior to the measurement.
}
\end{figure*}
\begin{figure*}[th]
\begin{center}
 \includegraphics[width=16cm]{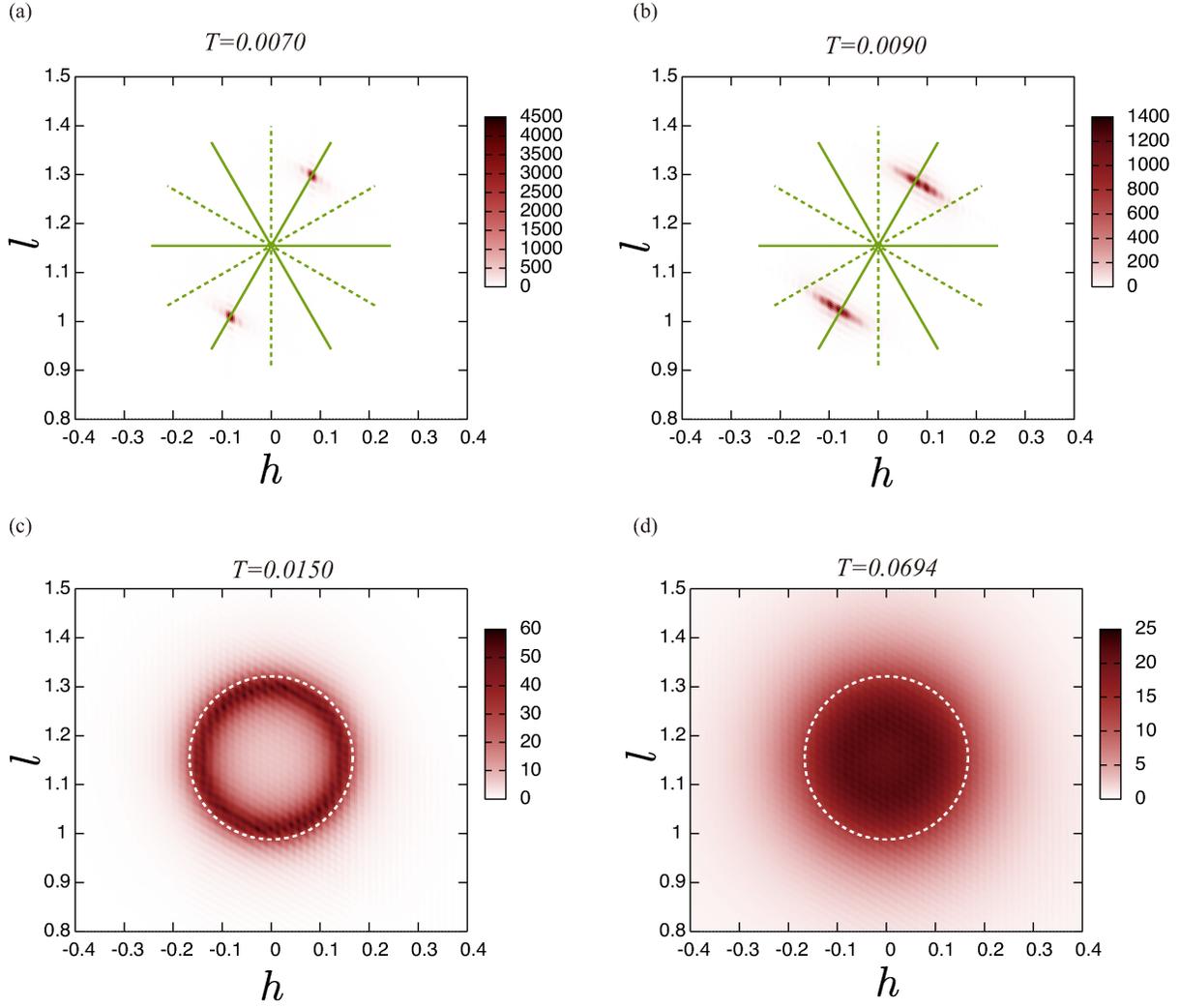}
\end{center}
 \caption{
(Color online) The intensity plot of the spin structure factor in the $(h,l)$ plane for $J_2/J_1=0.1830$, magnified for a part of the plane around an AF point $(0,2/\sqrt{3})$. The temperatures are $T=0.0070$ (a),  $T=0.0090$ (b),  $T=0.0150$ (c), and $T=0.0694$ (d), respectively. The lattice size is $L=72$. We turn off the temperature-exchange process when measuring $F(\bm{q})$ to probe the way of the symmetry breaking clearly, while it is turned on in equilibrating the system prior to the measurement. The next-nearest-neighbor (nearest-neighbor) directions of the honeycomb lattice are given by the green solid (broken) lines in (a) and (b). The ring corresponding to the ground-state degeneracy is shown by the broken circles in (c) and (d).
}
\end{figure*}
 At a temperature between $T^*$ and $T_{c}$, $F(\bm{q})$ still selects the state in the nnn direction, but now the intensity becomes {\it diffuse\/} exhibiting a ridge along the ring direction: See Fig.6(b). It means that the fluctuation in the $\bm{q}$-vector direction becomes enhanced.  At temperatures above $T_{c}$, the $Z_3$ symmetry is restored and $F(\bm{q})$ exhibits a pronounced ring-like pattern surrounding the AF point, which faithfully reflects the ground-state degeneracy: See Fig.6(c). On further increasing the temperature, the center of the ring is `buried', giving rise to a ``pancake-like'' pattern shown in Fig.6(d). Note that the radius of such a ``pancake'' is still determined by the radius of the degenerate ring, and hence, hardly depends on the temperature.

 In Fig.7(a), we show for $j_2=0.1830$ a segment of $F(\bm{q})$ along the $h=0$ axis at various temperatures. $F(\bm{q})$ exhibits the double-peak pattern corresponding to the ring for $0.013 \lsim T \lsim 0.06$ and the flat pancake-like pattern for $0.06 \lsim T \lsim 0.11$. Note that the pancake pattern is not fittable by the double-Lorenzian centered at the ring positions. These states with characteristic shapes of $F(\bm{q})$ are all paramagnetic states in the sense that no symmetry breaking occurs there, but they differ from the standard paramagnetic state in the sense that the associated $F(\bm{q})$ exhibits quite unusual shape. We call these exotic paramagnetic states a ``ring-liquid'' and a ``pancake-liquid'' state.

%
\begin{figure}
\begin{center}
 \includegraphics[width=8cm]{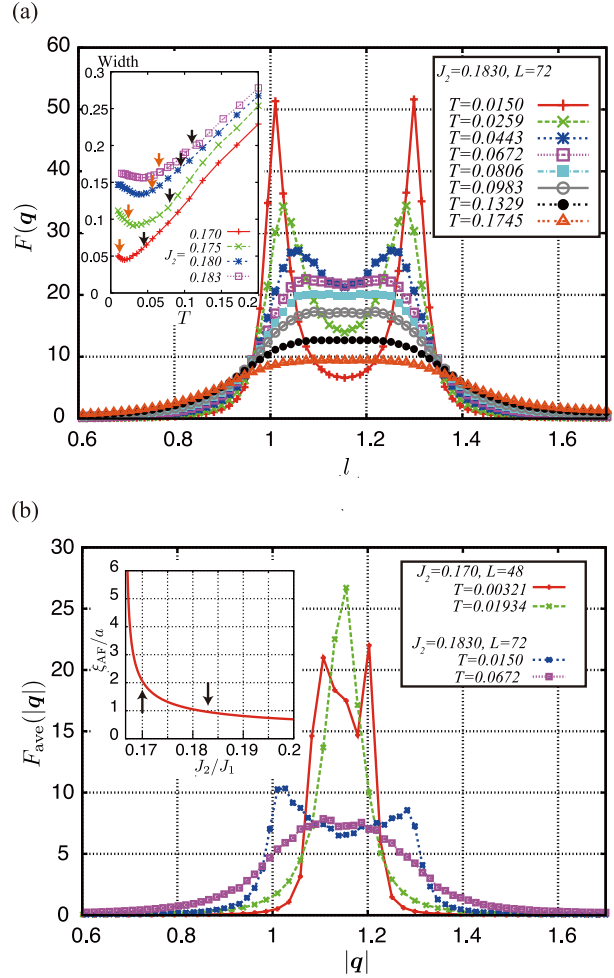}
\end{center}
 \caption{
(Color online) The spin structure factor in the ``ring-liquid'' and the
 ``pancake-liquid'' regions. (a) The $l$-dependence of the spin structure factor along the $h=0$ line for various temperatures above $T_c$ at $J_2/J_1=0.1830$. The lattice size is $L=72$. The ``ring-liquid'' state is realized at $0.013 \lsim T \lsim 0.06$, while the ``pancake-liquid'' state is realized at $0.06 \lsim T\lsim 0.11$. The inset represents the temperature dependence of the width (at half maximum) of the $F_{{\rm ave}}(|\bm{q}|)$-peak for several $J_2/J_1$-values. The crossover points between the ring-liquid and the pancake-liquid, and the ones between the pancake-liquid and the standard paramagnet are given by arrows. (b) The $|\bm{q}|$-dependence of the powder-averaged spin structure factor, {\it i.e.\/}, the spin structure factor averaged over angles with the modulus $|\bm{q}|$, in the ring-liquid and the pancake-liquid regions for $J_2/J_1=0.17$ and 0.1830.  The lattice size is $L=72$. The inset exhibits the $J_2/J_1$-dependence of the `antiferromagnetic correlation length' $\xi_{AF}$ in units of $a$ corresponding to the inverse width of the $F_{{\rm ave}}(|\bm{q}|)$-peak, which is given by the radius of the degenerate ring.
}
\end{figure}

 In principle, it should be possible to experimentally observe such characteristic changes of $F(\bm{q})$ by means of neutron scattering. Since only powder samples have been available so far for BMNO, we also calculate the powder averaged $F(\bm{q})$ for several typical cases, and the results are shown in Fig.7(b). The power-averaged $F(\bm{q})$ exhibits a peak centered around the AF $|\bm{q}|$-point with the width equal to the radius of the degenerate ring of the ground-state manifold.  Our present $F(\bm{q})$ seems fully consistent with the recent experimental data on BMNO \cite{Matsuda}. In our model calculation, the width of the peak in the ``ring-liquid'' or the ``pancake-liquid'' regimes, which would be interpreted as the inverse `AF correlation length' $\xi_{AF}$ experimentally, remains essentially temperature-independent. This is demonstrated in the inset of Fig.7(a) where the temperature dependence of the width of the $F_{{\rm ave}}(|\bm{q}|)$-peak is shown for several $J_2/J_1$-values. In the ring and the pancake regimes, the width exhibits negligible or even opposite temperature dependence. In the inset of Fig.7(b), we show the $J_2/J_1$-dependence of the `AF correlation length' $\xi_{AF}$ : Towards the AF boundary $j_{2c}=1/6$, $\xi_{AF}$ diverges as $\propto 1/\sqrt{j_2 - j_{2c}}$. Nevertheless, it stays short even fairly close to the AF phase boundary: For example, $\xi_{AF}\simeq a$ at $j_2=0.1830$, and  $\xi_{AF}\simeq 2a$ at $j_2=0.17$. The recent experiment on BMNO yields $\xi_{AF}\simeq 8{\rm A} \simeq 1.6a$ \cite{Matsuda}.

 Now, we touch upon the issue of $T^*$, a transition-like temperature manifesting itself in the lower specific-heat peak and in the sharp rise of the directional order parameter $m_3$. We deduce that $T^*$ corresponds to the $Z_2$-vortex binding-unbinding transition. It has been known that the frustration-induced noncollinear order of the Heisenberg spin sustains a characteristic topological excitation called a $Z_2$ vortex \cite{KawamuraMiyashita}. Indeed, recent studies have indicated that a topological transition driven by the binding-unbinding of such $Z_2$ vortices occurs in certain frustrated 2D Heisenberg models with the noncollinear spin order \cite{KawamuraYamamoto,Domenge}. Our preliminary MC study of the $Z_2$-vortex distribution supports such an identification. Further details will be reported elsewhere.

 Next, we study the effect of applied magnetic fields. In the vicinity
 of the AF phase boundary, fields tend to induce the AF order. This is
 demonstrated in Fig.8(a) where the AF order parameter $m_{AF}$ is
 plotted versus the field intensity for several temperatures at
 $J_2/J_1=0.175$ close to the AF phase boundary. The AF order parameter
 is defined by 
\begin{equation}
m_{AF}=\frac{1}{2}\langle
 |(\bm{m}_A^{xy}-\bm{m}_B^{xy})|\rangle , 
\end{equation}
where $\bm{m}_A^{xy}$ and  $\bm{m}_B^{xy}$ are the transverse components (perpendicular to the
 applied magnetic field) of the sublattice magnetization per spin of the
 sublattices A and B. One can see that the AF correlation is significantly enhanced by fields in the temperature range above $T_{c}$. Note that $m_{AF}$ should vanish in the thermodynamic limit in the 2D Heisenberg model, yet gives a measure of the AF correlations in finite systems. In real systems, a weak 3D coupling might eventually realize the true AF long-range order. The field dependence of the spin structure factor is also shown in Fig.8(b). The region of such a field-induced AF is also depicted in Fig.9(b), which occupies the region around the pancake-liquid regime. It would be natural to expect that the pancake-liquid, which may be regarded as a modified ring-liquid state where the AF component is enforced, is favored in the AF-state formation. We note that a similar field-induced AF has recently been reported on BMNO by Matsuda {\it et al\/} \cite{Matsuda}.
\begin{figure}
\begin{center}
\includegraphics[width=8cm]{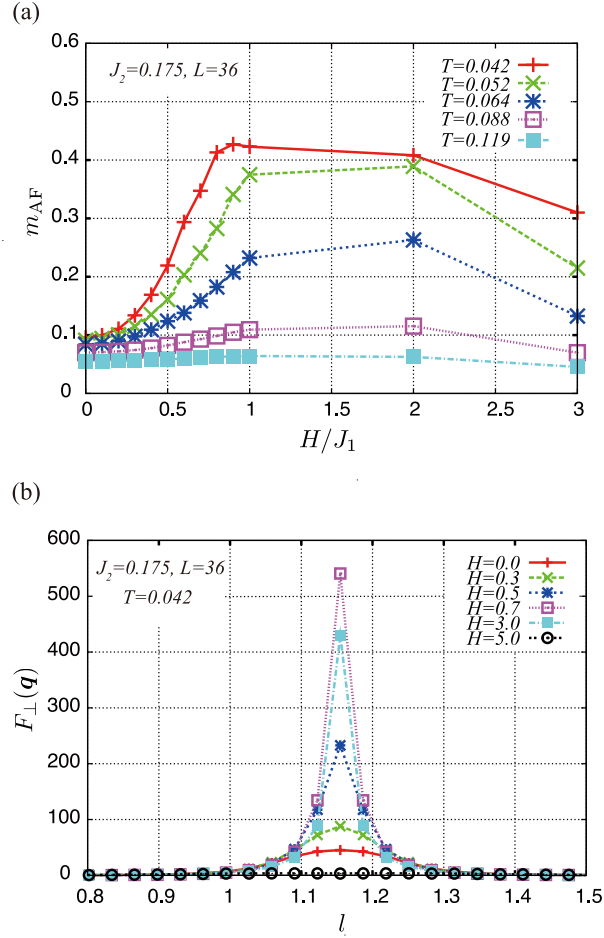}
\end{center}
\caption{
(Color online) (a) The magnetic field dependence of the AF order
 parameter $m_{AF}$. (b) The field dependence of the spin structure factor of the transverse components versus $l$ with $h=0$ for $J_2/J_1=0.175$. The temperature is $T=0.042$ above $T_c$. The lattice size is $L=36$. Applied fields induce a sharp peak around the AF point, $l=2/\sqrt{3}$.}
\end{figure}

 Our results are summarized in the $J_2/J_1$ versus temperature phase diagram of Fig.9(a). In Fig.9(b), the vicinity of the AF boundary is magnified by extending the temperature range to higher temperatures. Toward the AF phase boundary, the transition line tends to $T=0$, stabilizing the spin-liquid state without any symmetry breaking down to very low temperature. Indeed, $T_{c}/|T_{CW}|$ is estimated to be as low as $\sim 0.008$ at $j_2=0.1830$ and $\sim 0.001$ at $j_2=0.17$. In the vicinity of the AF phase boundary, the paramagnetic state is of unusual type characterized by the ring-like or the pancake-like spin structure factors, the ``ring-liquid'' or the ``pancake-liquid'' state.

%
\begin{figure}
\begin{center}
\includegraphics[width=8cm]{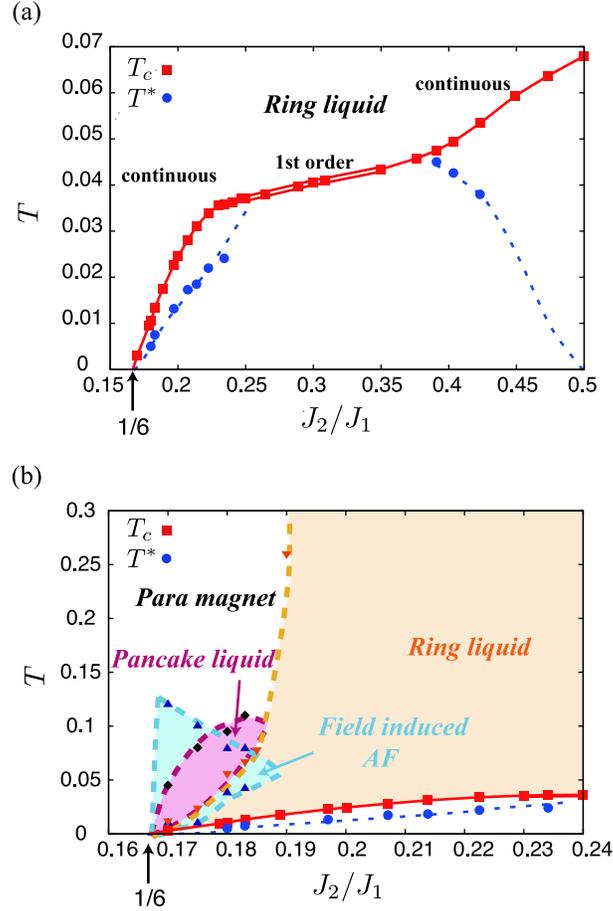}
\end{center}
\caption{
(Color online) Phase diagram of the model.
 (a) The $T_c$-curve representing the $Z_3$ directional-symmetry-breaking transition is plotted versus $J_2/J_1$, together with the $T^*$-curve determined from the lowest specific-heat-peak position for the largest size studied. Note that $T^*$ still depends on the system size appreciably so that the precise location of the $T^*$-curve is uncertain. We deduce that $T^*$ corresponds to the $Z_2$-vortex binding-unbinding transition (see the text). The directional-symmetry-breaking transition at $T_c$ is continuous in the smaller and in the larger $J_2/J_1$ regimes, but is first-order in the intermediate $J_2/J_1$ regime. There is a tendency that the directional-symmetry-breaking transition, which is originally continuous, becomes of first-order when it merges with the $T^*$-line. (b) represents the vicinity of the AF phase boundary $J_2/J_1=\frac{1}{6}$ including the higher temperature range above $T_c$. The regions of the ``pancake-liquid'', the ``ring-liquid'' and the standard paramagnetic states are indicated, each of which is separated by a crossover line. The crossover line between the ring-liquid and the pancake liquid is determined from the criterion whether $F(\bm{q})$ exhibits a single peak or double peaks, while that between the ring-liquid and the standard paramagnet is determined from the temperature dependence of the width of the $F_{{\rm ave}}(\bm{q})$-peak given in the inset of Fig.7(a). The region exhibiting the field-induced AF is indicated by shaded area, which is determined from the criterion that the maximum value of $m_{AF}$ induced by applied fields is greater than that in zero field by more than 50\%.
}
\end{figure}

 It should be emphasized that the strong frustration effect and the subsequent massive degeneracy alone are not enough to stabilize such spin-liquid states down to low temperature, since the order-by-disorder mechanism usually comes into play leading to the selection of a particular ordered state at the energy scale of the main exchange interaction $O(J_1)$. In the vicinity of the AF phase boundary of this model, the AF state, which remains to be a locally unstable state in itself, conspires with the frustration-induced massive degeneracy to stabilize the special types of spin-liquid states down to extremely low temperature. The ring-like degeneracy in the $\bm{q}$-space also plays an essential role in realizing novel spin-liquid states here, which is absent in, {\it e.g.\/}, the $J_1$-$J_2$ Heisenberg model on a square lattice, another 2D spin system exhibiting the order-by-disorder phenomena \cite{Weber}.

 If the $J_2/J_1$-value is increased away from the AF phase boundary,  $T_{c}$ increases, whereas the order of the transition changes from continuous to first-order at around $j_2\simeq 0.25$. It remains first-order for $0.25\lsim j_2\lsim 0.35$ and  becomes continuous  again beyond $j_2\gsim 0.35$: See Fig.9(a). Similar first-order $Z_3$ directional-symmetry-breaking transition was reported also for certain triangular-lattice Heisenberg AF \cite{Kawashima}.
The behavior of the $T_{c}$-line is consistent with the recent calculation by Mulder {\it et al\/} \cite{Mulder}, whereas the change in the order of the transition or the existence of another $T^*$-line was not reported there.

\section{Discussion and summary }
We wish to emphasize that our results are consistent with the
 recent experiment on BMNO \cite{Azuma, ESR, Matsuda} if BMNO lies close to the AF phase boundary. (i) The spin-liquid-like state is stabilized down to very low temperature. Our model provides a mechanism to significantly suppress the ordering in the vicinity of the AF phase boundary. (ii) Neutron-scattering intensity observed for powder samples, which exhibits an extremely broad peak centered around the AF point with apparently short AF correlation length, is consistent with the corresponding pattern expected for the pancake-liquid (or the ring-liquid) state of the present model. (iii) Experimentally observed field-induced AF is fully consistent with the corresponding phenomena observed in the pancake-liquid (or in the ring-liquid) state of the present model. (iv) Overall behavior of the specific heat and the susceptibility above $T_c$ are consistent with the experimental results \cite{Azuma}.

 Of course, in real experiments, several points not taken into account in the present model might play a role. Here we discuss the following two points. (a) Real BMNO is a bilayer honeycomb system with the AF interlayer coupling. We have also performed a preliminary simulation on the bilayer model, to find that the main results, particularly the points (i) $\sim $ (iv) above, do not change, at least qualitatively. (b) Real samples are likely to possess certain amounts of defects and impurities which might cause the glassy behavior at low temperatures, masking an intrinsic behavior of the pure system. For example, the expected phase transition has not been observed so far for BMNO down to 0.4K. The reason of this might be either (i) actual $T_c$ is lower than 0.4K, or (ii) the system at low temperatures is gradually stuck into the glassy state because of the randomness. In any case, further measurements on purer, and possibly, single crystal is desirable to clarify the issue.

 In summary, we have revealed an intriguing ordering behavior of the frustrated honeycomb-lattice Heisenberg AF. Near the AF phase boundary, exotic spin-liquid states like the ring-liquid or the pancake-liquid state are stabilized down to extremely low temperatures, accompanied by the characteristic spin structure factor and the field-induced antiferromagnetism. Our results seem consistent with the recent experimental data on BMNO.

\section*{Acknowledgements}
The authors are thankful to M. Azuma and M. Matsuda for useful discussion.
This work is supported by Grand-in-Aid for Scientific Research on Priority Areas ``Novel State of Matter Induced by Frustration'' (19052006 \& 19052008). We thank the Supercomputer Center, Institute for Solid State Physics, University of Tokyo for providing us with the CPUtime.

\appendix
\section{Low-temperature expansion}

Here we calculate the partition function ${\cal Z}$ of the model described by the Hamiltonian ${\cal H}$,
\begin{eqnarray}
 {\cal Z}&=&\int D\bm{S} e^{-\beta {\cal H}} \prod_{i} \delta[\bm{S}_{i}^{2}-1]
 \label{bunpai}
\end{eqnarray}
by means of a low-temperature expansion from an arbitrary state in the ground-state manifold. Let the ground-state spin orientation be $\bm{{\overline S}}_{i}$, which is assumed to lie in the $xy$ plane. $\bm{{\overline S}}_{i}$ is an incommensurate spiral state characterized by the wavevector $\bm{q}$ belonging to the degenerate ring of the ground-state manifold. By introducing the deviation vector $\bm{\pi}_{i}$, which satisfies $\bm{\pi}_{i}\perp \bm{{\overline S}}_{i}$, one has
\begin{center}
 $\displaystyle \bm{S}_{i}=\bm{\pi}_{i}+\bm{{\overline S}}_{i} \sqrt{1-\bm{\pi}_{i}^{2}}$ .
\end{center}

By decomposing the $\bm{\pi}_{i}$-vector into the $xy$- and $z$-components,
\begin{center}
 $\bm{\pi}_{i}= \bm{e}_{z}\phi_{i} +[\bm{e}_{z}\times
 \bm{{\overline S}_{i}}]\chi_{i}$
\end{center}
and expanding the Hamiltonian up to the quadratic order both in $\bm{\pi}$ and $\phi$, the partition function ${\cal Z}$
\begin{eqnarray}
 \displaystyle {\cal Z}&=&\int D\phi D\chi  e^{-\beta {\cal H}} \label{sayou}
\end{eqnarray}
can be evaluated by the Gaussian integrals. Neglecting the terms independent of the wavevector $\bm{q}$ specifying the ground state, which even include the divergent terms, we finally get the following expression of the $\bm{q}$-dependent part of the free energy,
\begin{multline}
F(\bm{q})/T =  \int d{\bm{q'}} (\ln[W_{11}(\bm{q,q'}) + |W_{12}(\bm{q,q'})|] \\
 + \ln[W_{11}(\bm{q,q'}) - |W_{12}(\bm{q,q'})|]) .
\end{multline}
where
\begin{multline}
W_{11}(\bm{q,q'})
 = 2J_{2}\Biggl[\cos q'_{x} \cos q_{x}\\
 + \cos \frac{q'_x+\sqrt{3}q'_y}{2} \cos \frac{q_x+\sqrt{3}q_y}{2}\\
 + \cos \frac{q'_x-\sqrt{3}q'_y}{2} \cos \frac{q_x-\sqrt{3}q_y}{2}\Biggr]
 - \lambda(\bm{q}),
\end{multline}
\begin{multline}
W_{12}(\bm{q,q'})
 = J_{1}\Biggl[ \cos \alpha(\bm{q}) e^{-i\frac{(q'_x+q'_y/\sqrt{3})}{2}} \\+
 \cos (\alpha(\bm{q})-\frac{q_x+\sqrt{3}q_y}{2})
 e^{iq'_y/\sqrt{3}} \\+ \cos
 (\alpha(\bm{q})-q_x)e^{i\frac{q'_x-q'_y/\sqrt{3}}{2}}\Biggr] .
\end{multline}
The function $\lambda(\bm{q})$ is defined by
\begin{equation}
\displaystyle \lambda(\bm{q}) =2J_{2}\epsilon(\bm{q}) - J_{1}\sqrt{3+2\epsilon(\bm{q})} ,
\label{lambda1}
\end{equation}
with
\begin{equation}
 \epsilon(\bm{q}) = \cos q_x+ \cos \frac{q_x+\sqrt{3}q_y}{2} + \cos \frac{q_x-\sqrt{3}q_y}{2} ,
\end{equation}
while $\alpha(\bm{q})$, representing a phase angle between the two nn spins in a unit cell belonging to the two sublattices A and B as $\theta_i^{(A)}=\bm{q}\cdot \bm{r}_i^{(A)}$ and $\theta_{i+\delta}^{(B)}=\bm{q}\cdot \bm{r}_i^{(A)}+\alpha(\bm{q})$, is given by
\begin{equation}
 \cos \alpha (\bm{q})= -(1+\cos q_x+\cos \frac{q_x+\sqrt{3}q_y}{2})/\sqrt{3+2\epsilon(\bm{q})} , \label{defalpha1}
\end{equation}
\begin{equation}
 \sin \alpha (\bm{q})= -(\sin q_x+\sin \frac{q_x+\sqrt{3}q_y}{2})/\sqrt{3+2\epsilon(\bm{q})} . \label{defalpha2}
\end{equation}


\begin{thebibliography}{99}

\bibitem{Anderson} P.W. Anderson: Mater. Res. Bull. {\bf 8} (1973) 153.

\bibitem{Einarsson} T. Einarsson and H. Johannesson: Phys. Rev. B{\bf 43} (1991) 5867.

\bibitem{Takano} K. Takano: Phys. Rev. B{\bf 74} (2006) 140402(R).

\bibitem{Kitaev} A. Kitaev: Ann. Phys. (N.Y.) {\bf 321} (2006) 2.

\bibitem{Kivelson} H. Yao and S.A. Kivelson: Phys. Rev. Lett. {\bf 99}
       (2007) 247203.
       
\bibitem{Cabra} D.C. Cabra, C.A. Lamas, and H.D. Rosales: arXiv:1003.3226.

\bibitem{Kane} C.L. Kane and E.J. Mele: Phys. Rev. Lett. {\bf 95} (2005) 146802; 226801.
\bibitem{Rachel} S. Rachel and K.L. Hur: arXiv:1003.2238.

\bibitem{Katsura} S. Katsura, T. Ide, and , T. Morita:
       J. Stat. Phys. {\bf 42}  (1986) 381.

\bibitem{Azuma} O. Smirnova, M. Azuma, N. Kumada, Y. Kusano, M. Matsuda,
       Y. Shimakawa, T. Takei, Y. Yonesaki, and N. Kinomura:
       J. Am. Chem. Soc. {\bf 131} (2009) 8313 .

\bibitem{Matsuda} M. Matsuda, M. Azuma, M. Tokunaga, Y. Shimakawa, and N
       Kumada: preprint.

\bibitem{ESR} S. Okubo, F. Elmasry, W. Zhang, M. Fujisawa, T. Sakurai,
       H. Ohta, M. Azuma, O. A. Sumirnova, and N. Kumada:
       J. Phys.:Conf. Ser. {\bf 200} (2010) 022042.
       
\bibitem{Villain} J. Villain, R. Bidaux, J.P. Carton, and R. Conte:
       J. Physique {\bf 41} (1980)  1263.

\bibitem{Kawamura} H. Kawamura: J. Phys. Soc. Jpn. {\bf 53} (1984) 2452.

\bibitem{Henley} C.L. Henley: Phys. Rev. Lett. {\bf 62} (1989) 2056.

\bibitem{Balents} D. Bergman, J. Alicea, E. Gull, S. Trebst and
       L. Balents: Nature Physics {\bf 3} (2007) 487.

\bibitem{KM85} H. Kawamura, and S. Miyashita: J. Phys. Soc. Jpn. {\bf
       54} (1985) 4530.

\bibitem{Zhitomirsky} M.E. Zhitomirsky: Phys. Rev. Lett. {\bf 88} (2002)
       057204.

\bibitem{Mulder} A. Mulder, R. Ganesh, L. Capriotti and A. Paramekanti:
       arXiv:1004.1119.

\bibitem{KawamuraMiyashita} H. Kawamura, and S. Miyashita:
       J. Phys. Soc. Jpn. {\bf 53} (1984) 4138.

\bibitem{KawamuraYamamoto} H. Kawamura, A. Yamamoto, and T. Okubo:
       J. Phys. Soc. Jpn. {\bf 76}  (2010) 023701.

\bibitem{Domenge} J.-C. Domenge, C. Lhullier, L. Messio, L. Pierre, and
       P. Viot: Phys. Rev. B {\bf 77} (2008) 172413.

\bibitem{Weber} C. Weber, L.Capriotti, G. Misguich, F. Becca,
       M. Elhajal, and F. Mila: Phys. Rev. Lett. {\bf 91} (2003) 177202.

\bibitem{Kawashima} R. Tamura, and N. Kawashima: J. Phys. Soc. Jpn. {\bf
       77} (2008) 103002.




\end{thebibliography}
\end{document}